\documentclass[a4paper,12pt]{article}
\usepackage[latin1]{inputenc}
\usepackage[T1]{fontenc}
\pdfoutput=1
\usepackage{amsmath}
\usepackage{amsfonts}
\usepackage{amssymb}
\usepackage[english]{babel}
\usepackage[pdftex]{graphicx,color}

\title{Dirichlet boundary conditions in a noncommutative theory}
\author{C.~D.~Fosco
and
P.~Scuracchio\\
{\normalsize\it Centro At\'omico Bariloche and Instituto Balseiro}\\
{\normalsize\it Comisi\'on Nacional de Energ\'\i a At\'omica}
\\
{\normalsize\it R8402AGP Bariloche, Argentina.}}
\begin{document}
\date{\today}
\maketitle
\begin{abstract}
We study the problem of imposing Dirichlet-like boundary conditions along
a static spatial curve, in a planar Noncommutative Quantum Field Theory model.

\noindent After constructing interaction terms that impose the boundary
conditions, we discuss their implementation at the level of an interacting
theory, with a focus on their physical consequences, and the symmetries they 
preserve.
We also derive the effect they have on certain observables, like the
Casimir energies.
\end{abstract}
\newpage
\section{Introduction}\label{sec:intro}
Casimir and related effects~\cite{intro} may be thought of as observable
results of the interplay between the geometrical properties of a certain
spatial region ${\mathcal M}$, subject to some conditions imposed on its
boundary $\partial {\mathcal M}$, and the vacuum fluctuations (inside and
outside ${\mathcal M}$) of a quantum field that experiences those boundary
conditions.  One of the best-known observable consequences of this
interplay is the Casimir force, whose properties depend on the details of
the system considered, like the nature of the boundary conditions to be
imposed, the number of spatial dimensions, and the kinematic properties of
the vacuum field itself (like its spin and internal space structure). 

A very special case occurs when the vacuum field is described by a
Noncommutative Quantum Field Theory (NCQFT)~\cite{Douglas:2001ba}, since
this single quality simultaneously affects all of the properties 
which observable effects may depend upon.  Indeed, a first, immediate fact, is
that
the spacetime structure where one should define ${\mathcal M}$ and
$\partial {\mathcal M}$, is different to the one of a commutative theory,
since the coordinates $x_\mu$ are noncommutative objects. In this respect,
the most frequently considered situation corresponds to assuming that, in
$d+1$ dimensions, the coordinates satisfy a relation that falls under the
general form: 
\begin{equation}\label{eq:nccoord} 
[x_\mu \,,\, x_\nu ] \;=\; i \, \theta_{\mu\nu} \;\;, \;\;\;\mu,\,\nu =
	\,0,\,1,\,\ldots,\,d\; \;\;\;, 
\end{equation} 
where $\theta_{\mu\nu}$ is a constant antisymmetric tensor.  
At the same time, the field is now an element in the noncommutative algebra
generated by the coordinates; thus, one should expect departures from the
commutative case also coming from this `kinematic' aspect.

It is our purpose in this paper to explore some of the distinctive
characteristics of this kind of system, namely, NCQFT with static,
Dirichlet-like boundary conditions. A very interesting phenomenological
approach to this kind of problem has been applied, in~\cite{Casadio:2007ec}, to
the Casimir effect for two parallel plates in $3+1$ dimensions.  We follow here
a different path, where the boundary region appears from the very beginning, and
the (unavoidable) `fuzziness' in the boundaries is automatically incorporated
(because of the noncommutativity) into the formulation. Besides, we attempt to
make sense of a boundary with a rather arbitrary shape. In this aspect,
we  generalize the work presented by one of us in~\cite{Fosco:2007tn}.  

We shall first deal with the formulation of the problem itself, since, as
we just mentioned, one needs to make sense of the noncommutative analogue
of geometrical regions, for example, $\partial{\mathcal M}$, something that
becomes problematic, because of the `uncertainty principle' which follows
from (\ref{eq:nccoord}).

An important ingredient in this study is the number of spacetime dimensions
of the system considered. We are dealing with {\em static\/} boundaries;
 with this in mind, we shall study theories where just the
spatial coordinates are noncommutative. As a bonus, we avoid
muddling the discussion with problems due to time-like noncommutativity,
which, although interesting, are not relevant to the kind of phenomenon we
want to study here. With this caveat, and restricting our scope to $d \leq 3$, the most 
interesting case certainly corresponds to $d=2$, namely, planar theories. Indeed,
if $d=1$, the theory would be commutative and thus irrelevant to our study. 
On the other hand, when $d=3$, only two spatial
coordinates can be noncommutative. Indeed, in the equation satisfied by the
spatial coordinates,
\begin{equation}
[x_j \,,\, x_k ] \;=\; i \, \theta_{jk} \;\;, \;\;\;j,\,k =
	\,1,\,\ldots,\,d\; \;\;\;, 
\end{equation} 
the antisymmetric tensor, being of odd order, has at least a zero
eigenvalue; i.e., there is a commutative coordinate.  Thus, to make the
geometry fully sensitive to noncommutativity, we shall consider a
$2+1$ dimensional model.  

Finally, there is also a phenomenological reason for
this choice: this situation is realized, in exactly that way, when a strong
constant magnetic field is applied to an essentially two-dimensional
system. A projection to the lowest Landau level then justifies a
noncommutative description~\cite{Dunne:1992ew,LLL}.  In this kind of planar
model, boundaries are known to play an important role, and one should
indeed expect it to be so because those systems are the quantum version of
incompressible fluids~\cite{Susskind:2001fb,Jackiw:2004nm}.

From a practical standpoint, since the time coordinate remains commutative,
the Hamiltonian plays the role of generator of infinitesimal time translations
in the usual way, hence many standard Quantum Field Theory tools retain the
usual interpretation they have in the commutative case.

The structure of this paper is as follows: in section~\ref{sec:model} we
define the model and introduce our notation and conventions, in
section~\ref{sec:boundary} we study the boundary conditions and discuss
their main properties. 
In~\ref{sec:self-dual}, we deal with the case of models including a
Grosse-Wulkenhaar term, at the self-dual point.
Section~\ref{sec:conc} contains our conclusions.
\section{The model}\label{sec:model}
We shall be concerned with a complex noncommutative scalar field $\phi$ in
$2+1$ dimensions, on which we shall define and impose (the noncommutative
version of) Dirichlet-like boundary conditions along a spatial curve
${\mathcal C}$ (see Figure 1), defined, for example, by its parametric
form: 
\begin{equation}\label{eq:param}
	{\mathcal C}\big) \;\;\; \xi  \, \longrightarrow \, z(\xi) \;,
\end{equation}
where $\xi$ is a real parameter. We shall use, throughout this paper, the
convention that spatial coordinates shall be denoted by letters from the
end of the Roman alphabet ($x$, $y$, $z$, \ldots), so that the
noncommutativity relations (which, by assumption, affect just those coordinates) may be
written as follows:
\begin{equation}
	[x_j \, , \, x_k ] \;=\; i \, \theta_{jk} \;\;, \;\;\theta_{jk} =
	\theta \; \epsilon_{jk} \;, \;\;\;\; j,\,k\,=\,1,\,2 \;.
\end{equation}
where $\theta$, the parameter controlling the strength of noncommutativity,
has the dimensions of an area.
The (commutative) Euclidean time coordinate, shall be denoted by $\tau$ or
$x_0$, depending on the context. 
\begin{figure}
	\begin{center}
	\begin{picture}(0,0)%
\includegraphics{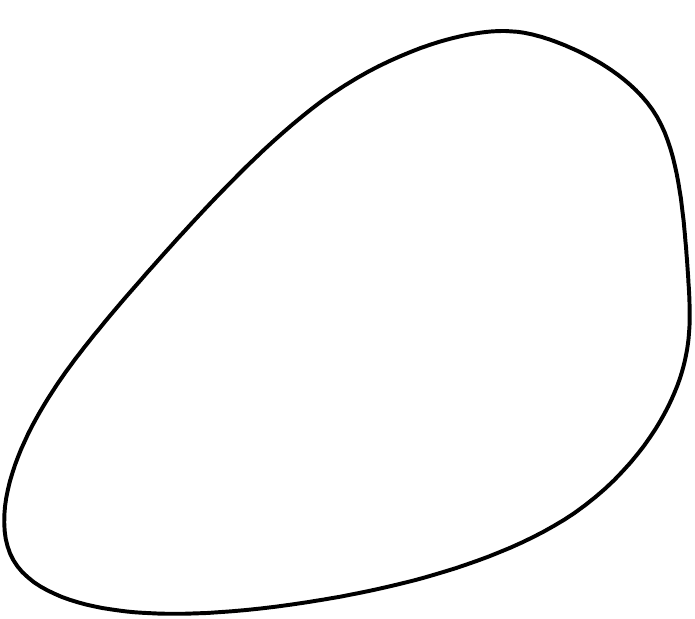}%
\end{picture}%
\setlength{\unitlength}{2486sp}%
\begingroup\makeatletter\ifx\SetFigFont\undefined%
\gdef\SetFigFont#1#2#3#4#5{%
  \reset@font\fontsize{#1}{#2pt}%
  \fontfamily{#3}\fontseries{#4}\fontshape{#5}%
  \selectfont}%
\fi\endgroup%
\begin{picture}(5288,4698)(3048,-6082)
\put(7021,-1591){\makebox(0,0)[lb]{\smash{{\SetFigFont{9}{10.8}{\rmdefault}{\mddefault}{\updefault}{\color[rgb]{0,0,0}${\mathbf{\mathcal C}}$}%
}}}}
\put(5086,-4156){\makebox(0,0)[lb]{\smash{{\SetFigFont{10}{12.0}{\rmdefault}{\mddefault}{\updefault}{\color[rgb]{0,0,0}${\mathbf{\mathcal M}}$}%
}}}}
\end{picture}%
	\caption{The region ${\mathcal M}$ and its boundary ${\mathcal C}$}
\end{center}
\end{figure}
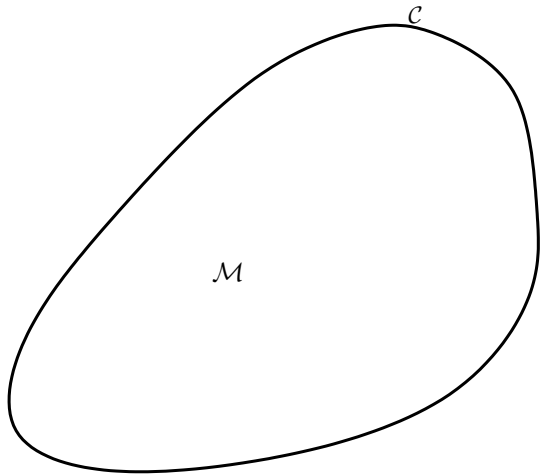
The curve ${\mathcal C}$ defined above is a standard, commutative geometry
object. To impose boundary conditions on the NCQFT, we follow the approach
of~\cite{Fosco:2007tn},
where Dirichlet-like conditions are imposed by means of a specific
interaction term which makes use of  the noncommutative version  of a
$\delta$-like interaction term. This will (as we shall see) induce some
special conditions for the fields on a `fuzzy' region determined by the curve,
at the same time preserving the property of having the right commutative
limit.

The scalar field $\phi$ will be equipped with an Euclidean action $S$, a
functional of the curve ${\mathcal C}$, the field $\phi$ and its adjoint
$\phi^\dagger$, with the following structure:
\begin{eqnarray}
	S\big({\mathcal C};\phi^\dagger,\phi \big) &=& S_0\big({\mathcal
	C};\phi^\dagger,\phi \big) +  
	S_I\big(\phi^\dagger,\phi \big) \nonumber\\
	S_0\big({\mathcal C};\phi^\dagger,\phi \big) &=&
	S_f\big(\phi^\dagger,\phi \big) \,+\,
	S_b\big({\mathcal C};\phi^\dagger,\phi \big) 
\end{eqnarray}
where $S_f$ denotes the action for a non-interacting field in the absence
of borders, $S_b$ is a term which accounts for the boundary conditions, and
$S_I$ corresponds to self-interactions. More explicitly, we shall assume
for $S_f$ the form:
\begin{equation}
	S_f\big(\phi^\dagger,\phi \big)\;=\; 
\int d^3x \,\big[ (\partial_\mu \phi)^\dagger \star \partial_\mu \phi \,+\,
m^2 \,\phi^\dagger \star \phi \big] 
\end{equation}
while for $S_b$, inspired by the commutative case, we consider two
non-equivalent forms, $S_b^{(L)}$ and
$S_b^{(R)}$ (both of which have the same $\theta \to 0$ limit): 
\begin{eqnarray}
	S_b^{(L)}\big(\phi^\dagger,\phi \big) &=& \lambda_L \,\int d^3x \,
	\phi(x) \star \delta_{\mathcal C}(x) \star \phi^\dagger(x)
\nonumber\\
	S_b^{(R)}\big(\phi^\dagger,\phi \big) &=& \lambda_R \,\int d^3x \,
	\phi^\dagger(x) \star \delta_{\mathcal C}(x) \star \phi(x)
\end{eqnarray}
where $\delta_{\mathcal C}(x)$ is the noncommutative analogue of 
the commutative two-dimensional $\delta_{\mathcal C}$ function which has support on
${\mathcal C}$. 
The effect of adding each one of the boundary terms above plays, shall be
elucidated at the end of next section.

In terms of the  parametric form for ${\mathcal
C}$, the Weyl mapping immediately yields:  
\begin{equation}\label{eq:delta1}
	\delta_{\mathcal C}(x) \;=\; \int d\xi \, |\dot{z}(\xi)|
	\, \delta^{(2)}\big[ x - z(\xi)\big]
\end{equation}
with $\dot{z}(\xi) \equiv \frac{d}{d\xi}z(\xi)$. Note that, in the
expression above (only) $x_1$ and $x_2$ are noncommutative objects.  It is
worth noting that, to make sense of $\delta^{(2)}$ above one makes the
implicit assumption that:
\begin{equation}\label{eq:delta2}
	\delta^{(2)}\big[ x - z(\xi)\big] \;=\; \int \frac{d^2k}{(2\pi)^2}
	\, e^{i k \cdot [ x - z(\xi)]} \;.
\end{equation}
It will in some cases be useful to consider an alternative implicit
representation of the curve. Indeed, assuming that ${\mathcal C}$ can also be
described by the equation ${\mathcal F}_{\mathcal C}(x) = 0$,
the {\em commutative\/} version of $\delta_{\mathcal C}(x)$ can be written
as follows: 
\begin{equation}\label{eq:delta3}
	\delta_{\mathcal C}(x) \;=\; 
\big|{\mathbf \nabla} {\mathcal F}_{\mathcal C}(x) \big| \;\; \delta\big[{\mathcal F}_{\mathcal C}(x) \big] \;.
\end{equation}
This representation, except for some special cases, is not suitable for
immediate translation to the noncommutative case, because of the possible
noncommutativity between the two factors in (\ref{eq:delta3}). 
Special cases where that problem does not show up are: 1) A linear
function, namely: ${\mathcal F}_{\mathcal C} = a_1 x_1 + a_2 x_2$, 
which corresponds to a straight line, and 
${\mathcal F}_{\mathcal C} = \sqrt{x_1^2 + x_2^2} - R$,
which yields a radius $R$ circle. 
In the general case, $|\nabla {\mathcal F}_{\mathcal C}|$ may fail to commute
with ${\mathcal F}_{\mathcal C}$ . However, except when the gradient vanishes on points
belonging to ${\mathcal C}$, one can in principle find an alternative function, 
${\mathcal F}'_{\mathcal C}$ say, such that its gradient does have unit modulus
on (${\mathcal C}$), yet defining the same curve. Indeed, one may just define:
\begin{equation}
{\mathcal F}'_{\mathcal C}(x) \,=\, g(x) \,{\mathcal F}_{\mathcal C}(x) \;
\end{equation} 
where $g$ is a function such that:
\begin{equation}
g(x)\equiv \big|{\mathbf \nabla}{\mathcal F}_{\mathcal C}(x)\big|^{-1}
\end{equation}
for every  $x \in {\mathcal C}$. When $x \not \! \in {\mathcal C}$,
$g$ is smoothly extended quite arbitrarily, except for the condition of
being positive. 

Thus, whenever we use the implicit representation, we shall assume that it
has been already `normalized', such that:
\begin{equation}
\delta_{\mathcal C}(x) \;=\;  \delta\big[{\mathcal F}_{\mathcal C}(x) \big] 
\end{equation}
which can be unambiguously translated to its noncommutative
version~\footnote{As
usual, we map functions to Weyl-ordered operators.}.
The implementation and meaning of the $S_b$ terms are discussed at length in
the next section.

Finally, regarding the interaction term, we shall assume the two
inequivalent possible ones that can be built out a (globally invariant
under $U(1)$ transformations) quartic monomial in the field and its adjoint:
\begin{equation}
S_I \;=\; S_I^{(1)} \,+\, S_I^{(2)} 
\end{equation}
where 
\begin{eqnarray}
S_I^{(1)} &=& g^{(1)} \, \int d\tau d^2x \, \big[\phi^\dagger(\tau,x)
\star \phi(\tau,x) \star \phi^\dagger(\tau,x) \star \phi(\tau,x)
\big]
\nonumber\\
S_I^{(2)} &=& g^{(2)} \, \int d\tau d^2x \,
\big[\phi^\dagger(\tau,x)
\star \phi^\dagger(\tau,x) \star \phi(\tau,x) \star \phi(\tau,x)
\big] \;,
\end{eqnarray}
which correspond, respectively, to  planar and nonplanar vertices in a
perturbative expansion.

We shall make use of the alternative `operatorial' representation
for noncommutative actions, where the fields are not expanded apriori in
any basis, since it is useful to prove some special properties. In that
representation one has, for example:
\begin{equation}
\int d^2x \, \phi^\dagger(x) \phi(x) \phi^\dagger(x) \phi(x) 
\;\leftrightarrow \; 2 \pi \theta \, {\rm Tr}\big[ \phi^\dagger \, \phi \;
\phi^\dagger \, \phi \big] \;,
\end{equation} 
 (where we have adopted, for the sake of simplicity, the convention to use
the same symbol for a field in the Moyal representation as in the operatorial
version).

We conclude this section by noting that {\em except for the boundary
terms\/} the action is invariant under
the discrete transformation $\phi \to \phi^\dagger$, $\phi^\dagger \to
\phi$, which amounts to a kind of charge conjugation symmetry.

\section{The nature of the boundary conditions}\label{sec:boundary}
Some general properties can be deduced from the inclusion of
the boundary interaction term, $S_b$, as defined by a `normalized' function
${\mathcal F}_{\mathcal C}$. It is quite clear that this term, in the
commutative limit, yields Dirichlet-like boundary conditions. It is our
purpose here to see its effect in the noncommutative case. 
The intuitive idea that noncommutativity introduces some `fuzziness' into the
game, and therefore the boundary conditions will affect the field on a
finite-width region can be made more concrete. Indeed, 
taking into account the relation exploited in~\cite{Mezincescu:2000zq}, we see that:
\begin{equation}
	S_b^{(R)} \;=\; \lambda_R \, \int d\tau d^2x \, \phi^\dagger(\tau,x) \,
	\delta_{\mathcal C} (x + i \frac{\theta}{2} \, \tilde{\partial}\, )
	\phi(x)
\end{equation}
with: $\tilde{\partial}_j \equiv \epsilon_{jk} \partial_k$.
Then, by expanding in powers of $\theta$: $S_b^{(R)} =
S_b^{(R)}\big|_{\theta=0} \,+\, \delta S_b^{(R)}$, where the first term
contains the purely commutative part, and:
\begin{eqnarray}
	\delta S_b^{(R)} &=& -\frac{i \lambda_R \,\theta}{2} \, \int d\tau
	d^2x \, \delta_{\mathcal C}(x) \epsilon_{jk} \partial_j
	\phi^\dagger(\tau,x) \partial_k \phi(\tau,x) \nonumber\\
	 &+& \frac{\lambda_R \, \theta^2}{8} \, \int d\tau
	 d^2x \; \delta_{\mathcal C}(x)  \, \partial_{j_1} \partial_{j_2} \phi^\dagger(\tau,x) 
	 ( \delta_{j_1j_2} \partial^2 - \partial_{j_1} \partial_{j_2} )
	 \phi(\tau,x)  \nonumber\\
	&+&  \ldots 
\end{eqnarray}
This means that, at least in an expansion in $\theta$, the boundary
condition shall involve not just the field value on the curve but 
also its values on points that are close to it. This will manifest in the
existence of an effective width. 

To understand the effect of this term, it is convenient to go back,
momentarily, to the operatorial representation of the action. 
In the $S_b^{(R)}$ case, we have: 
\begin{equation}
S_b^{(R)} \;=\; 2 \pi \theta \,\lambda_R\, \int d\tau \, {\rm Tr} 
\big[ \phi^\dagger(\tau,x) \delta_{\mathcal C}(x)  \phi(\tau,x) \big] \;,
\end{equation}
where $\delta_{\mathcal C}(x) = \delta[{\mathcal F}_{\mathcal C}(x)]$, and
a similar expression for $S_b^{(L)}$.

To proceed, it is convenient to introduce a suitable representation (a basis), for the
fields, such that the $\delta$-function becomes diagonal. Of course, that
is tantamount to ${\mathcal F}_{\mathcal C}$ (a self-adjoint operator)
itself being diagonal.  There is a qualitative difference between two
possible cases, according to the spectrum of that
function being continuous or discrete.  The reason is intimately linked to
a geometrical property of the (commutative) curve ${\mathcal C}$:
this curve is assumed to divide space into two (`inner' and `outer')
components: an ${\mathcal F}_{\mathcal C}$ with a continuous spectrum
corresponds, in the commutative limit, to a case where the two regions are
unbounded. A discrete one, on the other hand, amounts to cases where one of
the regions has finite area.

Indeed, the commutative limit may be obtained by assuming that $\theta$ is
very small in comparison with the area element. Thus, regarding 
${ \mathcal F}_{\mathcal C}$ as the `Hamiltonian' for a fictitious time
evolution, one can obtain its eigenvalues in a Bohr-Sommerfeld like
fashion.
Then the area becomes quantized (discrete spectrum) only for a curve that encircles
a finite area. 
Note that this remark yields a possible concrete way to construct the boundary
conditions for a given curve: look for a (one dimensional) quantum mechanical
Hamiltonian such that, in the semiclassical limit, has a constant energy curve
that coincides with the boundary one wants to consider. That Hamiltonian, minus
the corresponding energy, is essentially equal to ${\mathcal F}_{\mathcal C}$.
Of course, one can always use the parametric representation instead.

Simple examples of the two kinds of spectra are: 
${\mathcal F}_{\mathcal C}= a_1 x_1 + a_2 x_2$ (continuous spectrum) and 
${\mathcal F}_{\mathcal C} = \sqrt{x_1^2 + x_2^2} - R$ (discrete spectrum).
We shall discuss the effect of $S_b$ on the properties of the corresponding
model for those two cases separately below.
 
\subsection{Continuous spectrum}\label{ssec:contspec}
Let $\{ | \epsilon \rangle\}$ be an orthonormal basis for ${\mathcal
F}_{\mathcal C}$:
\begin{equation}
{\mathcal F}_{\mathcal C} \, |\epsilon \rangle \;=\; 
\epsilon  \, | \epsilon \rangle \;,\;\;\; \langle \epsilon'|\epsilon
\rangle \,=\, \delta(\epsilon' - \epsilon) \;.
\end{equation}
We then expand the field in this basis. Using a shorthand notation for the
integrals, the expansion amounts to:
\begin{equation}
\phi(\tau) = \int_{\epsilon,\epsilon'} \, \phi_{\epsilon,\epsilon'}(\tau) \,
|\epsilon \rangle \langle \epsilon'| 
\;,\;\;\;
\phi^\dagger(\tau) = \int_{\epsilon,\epsilon'} \, \phi^*_{\epsilon',\epsilon}(\tau) \,
|\epsilon \rangle \langle \epsilon'| \;.
\end{equation}
We then get for $S_b^{(R)}$ the following expression:
\begin{equation}
S_b^{(R)} \;=\; 2 \pi \theta \, \lambda_R \, \int d\tau \; \int_{\epsilon} \,
|\phi_{0,\epsilon}(\tau)|^2 \;, 
\end{equation}
which, in the Dirichlet ($\lambda \to \infty$) limit imposes the condition: 
\begin{equation}\label{eq:dirichlet1}
\phi_{0,\epsilon}(\tau) \;=\; \langle 0 |\phi(\tau)| \epsilon\rangle =
0 \;,\;\; \forall \epsilon 
\end{equation}
or, in terms of operators, $\langle 0 | \phi(\tau)= 0$ (and
$\phi^\dagger(\tau) | 0 \rangle = 0$~\footnote{We have adopted a bra-ket
like notation to denote matrix elements of  objects in the algebra.
Those should not be confused with Fock-space matrix elements of quantum
field operators.}).

On the other hand, an analogous procedure for the $S_b^{(L)}$ term, yields:
\begin{equation}
S_b^{(L)} \;=\; 2 \pi \theta \, \lambda_L \int d\tau \; \int_{\epsilon} \,
|\phi_{\epsilon, 0}(\tau)|^2 \;. 
\end{equation}
And, in the Dirichlet limit: 
\begin{equation}\label{eq:dirichlet2}
\phi_{\epsilon,0}(\tau) \;=\; \langle \epsilon |\phi(\tau)| 0\rangle =
0 \;,\;\; \forall \epsilon 
\end{equation}
or, in terms of operators, $\phi(\tau) |0 \rangle = 0$ (and
$\langle 0 | \phi^\dagger(\tau)  = 0$). 

The two boundary terms can also be written as, 
\begin{eqnarray}
	S_b^{(R)} &=& 2 \pi \theta \, \lambda_R \int d\tau 
	\,{\rm Tr}\big[\phi^\dagger(\tau) {\mathcal P}_{\mathcal C} \phi(\tau) \big]
	\nonumber\\ 
	S_b^{(L)} &=& 2 \pi \theta \, \lambda_L \int d\tau 
	\,{\rm Tr}\big[\phi(\tau) {\mathcal P}_{\mathcal C} \phi^\dagger(\tau) \big]
	\;,
\end{eqnarray}
with ${\mathcal P}_{\mathcal C} \equiv |0\rangle \langle 0|$. 

To see this kind of constraint at work, we consider a concrete example of a
continuous spectrum, or, as already explained, of an unbounded, open curve. 
The simplest possible case is a straight line: 
${\mathcal C} \equiv \{ (x_1,x_2): x_2 = 0\}$. 

For the sake of concreteness, we consider the case of $S_b^{(R)}$,
since $S_b^{(L)}$ may be obtained by performing
quite straightforward changes which we consider at the end of this section
(see below). 

Rather than using the operator formulation, we go directly to the Moyal
representation, to see how it affects the plane-wave solutions one could obtain from
considering just $S_f$.  In this situation, we have:
\begin{equation}
	S_b^{(R)}\big(\phi^\dagger,\phi \big)\;=\; \lambda_R \int d\tau \int
	d^2x \; \phi^\dagger(\tau,x) \star \delta(x_2) \star \phi(\tau,x) \;,
\end{equation}
which, by Fourier transforming with respect to the translation invariant
coordinates, $\tau$ and $x_1$, adopts the form:
\begin{equation}
	S_b^{(R)} \;=\; \lambda_R \int \frac{dk_0}{2\pi} \frac{dk_1}{2\pi}  \, 
	\widetilde{\phi}^*(k_0,k_1;-\frac{\theta k_1}{2})
	\,
	\widetilde{\phi}(k_0,k_1;-\frac{\theta k_1}{2}) \;.
\end{equation}
Thus, we find out the equations of motion that follow from
considering just the action defined by $S_0$, by performing the same Fourier
transformation on the $S_f$ term. In real-time:
\begin{equation}
(\partial_2^2 + k_2^2) \, {\widetilde\phi}(k;x_2)
	\;=\; \int dx'_2 \, {\widetilde V}(k;x_2,x'_2) \,
	{\widetilde\phi}(k;x'_2)
\end{equation}
where we have introduced the kernel:
\begin{equation}\label{eq:vk}
	{\widetilde V}(k;x_2,x'_2) \;=\;  \lambda \, \delta(x_2 +
	\frac{\theta k_1}{2}) \,  \delta(x'_2 +
	\frac{\theta k_1}{2}) \;,  
\end{equation}
and the mass-shell condition yields: $k_2 \equiv \sqrt{k_0^2 -k_1^2} > 0$,
where we use the positive square-root, to consider a solution which
corresponds to a plane wave incident from negative values of $x_2$.

This kind of solution is naturally treated as an scattering problem; hence
we may write the solution to this problem by means of the corresponding
Lippmann-Schwinger (L-S) integral equation:
\begin{eqnarray}\label{eq:ls1}
	{\widetilde\phi}(k;x_2) &=& 
	{\widetilde\phi}^{(0)}(k;x_2) \nonumber\\
	&+&\int dx'_2 \int dx''_2 \, \Delta_R(k_2;x_2,x'_2) 
	{\widetilde V}(k;x'_2,x''_2) {\widetilde \phi}(k,x''_2) \;,
\end{eqnarray}
where ${\widetilde\phi}^{(0)}$ is the (incident) free-particle
wave, solution of
\begin{equation}
	(\partial_2^2 + k_2^2) {\widetilde\phi}^{(0)} (k_2;x_2) = 0 \;.
\end{equation} 
and $\Delta_R$ is the retarded Green's function.

Since we assume the free-particle solution to correspond to a wave incident from $x_2 < 0$,
${\widetilde\phi}^{(0)}(k_2;x_2) = e^{i k_2 x_2}$. Besides, 
the retarded Green's function satisfies:
\begin{equation}
	(\partial_2^2 + k_2^2) \Delta_R(k_2;x_2 - x'_2)  = \delta(x_2 -x'_2)\;,
\end{equation}
(with retarded boundary conditions) and may be written more explicitly as follows:
\begin{equation}
	\Delta_R(k_2; x_2 - x'_2) \;=\; \frac{i}{2 k_2} \, e^{ i k_2 \, |x_2 - x'_2| }\;. 
\end{equation}
To solve the L-S equation, we proceed in an entirely analogous way to the
one presented in~\cite{Fosco:2009ic}, obtaining:
\begin{equation}\label{eq:lss}
	{\widetilde \phi}(k;x_2) \;=\; e^{i k_2 x_2} \,+\, 
r(k) e^{ - i \frac{\theta k_1 k_2}{2}} e^{ i k_2 |x_2 + \frac{\theta k_1}{2}|} \;,
\end{equation}
where: $r(k) \equiv -\frac{ \frac{i\lambda}{2 k_2}}{ 1 +\frac{i\lambda}{2
k_2}}$.
From the equation above we can extract, by considering the situations where $x_2>0$
and $x_2<0$, the properties of the transmitted and reflected
waves, respectively. In particular, one may study the dependence of the
transmission and reflection coefficients on the incident momentum; in this
respect, we recall that the wave also has an $x_1$ dependence, which has been
factored out of the solution (since it is unaffected by the boundary
condition). However, there remains a dependence  on $k_1$ in the
behaviour of the $x_2$-dependent part. 

For the transmitted wave, ${\widetilde\phi} \equiv {\widetilde\phi}_>$, and  we have:
\begin{equation}
	{\widetilde \phi}_>(k;x_2) \;=\; [ 1  \,+\, r(k) ] \, e^{ i k_2 x_2} \;,
\end{equation}
whenever $x_2 + \frac{\theta k_1}{2} > 0$, exactly as in the commutative
case. In particular, there is no transmitted wave when $\lambda \to \infty$
and $x_2 + \frac{\theta k_1}{2} > 0$, since $r(k) \to -1$.
Note, however, that there is an important qualitative difference for points
such that  $x_2 + \frac{\theta k_1}{2} < 0$ (and $x_2 > 0$). 
Here we have instead:
\begin{equation}
	{\widetilde \phi}_>(k;x_2) \;=\; e^{i k_2 x_2} \,+\, r(k) 
\,e^{- i \theta k_1 k_2}  e^{- i k_2 x_2} \;,
\end{equation}
or, in the Dirichlet limit:
\begin{equation}
	{\widetilde \phi}_>(k;x_2) \;=\; e^{i k_2 x_2} \,-\,e^{- i \theta k_1 k_2}  e^{- i k_2 x_2} \;,
\end{equation}
which is quite different to the previous case. In particular, the
transmitted current is smaller than for $x_2 + \frac{\theta k_1}{2} > 0$,
to the point of vanishing in the Dirichlet limit. There is a reflected wave,
and the system behaves as if
there were a reflecting wall at $x_2 = -\frac{\theta k_1}{2}$, which only
acts if $k_1 < 0$ (we assume $\theta >0$). 

The behaviour at $x_2<0$ is consistent with this picture. Indeed, if
$k_1<0$ (for the `wall' to act), we see that
\begin{equation}
	{\widetilde \phi}_<(k;x_2) \;=\; e^{i k_2 x_2} \,+\, r(k) 
\,e^{- i \theta k_1 k_2}  e^{- i k_2 x_2} \;,
\end{equation}
since, in this case, $x_2 + \frac{\theta k_1}{2} < 0$. We see that the
outcome for the reflected wave corresponds to the same result as in the
commutative case, except that (due to the extra $\theta$-dependent phase)
it corresponds to a wall at $x_2 = - \frac{\theta k_1}{2}>0$,
which introduces a spatial shift of twice that value, since the wave has to
go forward and bounce back.
Finally, also for negative values of $x_2$, if  $0 > x_2  > -\frac{\theta k_1}{2}$, which 
is nonempty for $k_1>0$, we have:  
\begin{equation}
	{\widetilde \phi}_<(k;x_2) \;=\;  [ 1 \,+\, r(k)]\, e^{ i k_2 x_2} \;,
\end{equation}
meaning that there is no reflected component: it only appears when  $x_2  <
-\frac{\theta k_1}{2}$.

We may summarize all of the above by the statement that if we imagine
sending a wave packet with a non vanishing momentum dispersion in the $x_1$
direction, the boundary behaves as it had a finite width $\delta x_2 \sim
\theta \delta k_1$. Or, by a standard application of the usual uncertainty
principle for $\delta x_1$ and $\delta x_2$: 
\begin{equation}
\delta x_1 \; \delta x_2 \, \sim \, \theta \;.
\end{equation} 

In spite of this expected effect, note that,  for a single wall one can impose Dirichlet-like
boundary conditions. The resulting reflection and transmission coefficients
coincide with their commutative counterparts, but the wall will have an
effective momentum ($k_1$) dependent position.

As a consistency check on the previous derivation, we note that
condition (\ref{eq:dirichlet1}) can be translated into the Fourier
transformed fields of the Moyal representation as follows:
\begin{equation}\label{eq:dirichlet3}
0 \;=\; \int \frac{d^2k}{(2\pi)^2} \, \langle 0| e^{i k \cdot x}
|x_2\rangle \; \widetilde{\phi}(k) \;\;\;,\;\;\;\; \forall x_2 \;;
\end{equation}
or, more explicitly,
\begin{equation}
\widetilde{\phi}(-\frac{x_2}{\theta} ; \frac{x_2}{2}) \;=\; 0  \;\;\;,
\;\;\;\;
\forall x_2 \;,
\end{equation}
(in the same hybrid Fourier representation we used above). This is of
course equivalent to: 
\begin{equation}
\widetilde{\phi}(k_1 ; -\frac{\theta k_1}{2}) \;=\; 0  \;\;\;,
\;\;\;\;
\forall k_1 \;,
\end{equation}
as it should be.

A similar derivation for the $S_b^{(L)}$ term:
\begin{equation}
	S_b^{(L)}\big(\phi^\dagger,\phi \big)\;=\; \lambda \, \int d\tau \int
	d^2x \; \phi(\tau,x) \star \delta(x_2) \star \phi^\dagger(\tau,x) \;,
\end{equation}
after the same Fourier transformation yields:
\begin{equation}
	S_b^{(L)} \;=\; \lambda \, \int \frac{dk_0}{2\pi} \frac{dk_1}{2\pi}  \, 
	\widetilde{\phi}^*(k_0,k_1; \frac{\theta k_1}{2})
	\widetilde{\phi}(k_0,k_1; \frac{\theta k_1}{2})
\end{equation}
which imposes, in the Dirichlet limit, the condition:
$\widetilde{\phi}(k_0,k_1; \frac{\theta k_1}{2}) =0$, $\forall k_1$.

A simple extension of the single straight-line case is that of two parallel
straight lines, at $x_2=0$ and $x_2=l$. Since the boundary has two disjoint
components, one can consider different choices regarding the $L$ and $R$
boundary terms on each mirror. 

In the $RR$ case, the resulting boundary interaction term becomes:
\begin{eqnarray}
	S_b^{(RR)} &=& \lambda \, \int \frac{dk_0}{2\pi} \frac{dk_1}{2\pi}  \, 
\big[
\widetilde{\phi}^*(k_0,k_1;-\frac{\theta k_1}{2})
\,
\widetilde{\phi}(k_0,k_1;-\frac{\theta k_1}{2}) \nonumber\\
&+&
\widetilde{\phi}^*(k_0,k_1; l-\frac{\theta k_1}{2})
\,
\widetilde{\phi}(k_0,k_1;l-\frac{\theta k_1}{2}) 
\big] 	\;.
\end{eqnarray}
In particular, when $\lambda \to \infty$, we find that ${\widetilde \phi}$
has to satisfy the conditions $\widetilde{\phi}(k_0,k_1;-\frac{\theta
k_1}{2})=0$ and  $\widetilde{\phi}(k_0,k_1;l-\frac{\theta k_1}{2})=0$. The
solution is nontrivial if $k_2= \frac{n \pi}{l}$, exactly as in the
commutative case.  Thus the Casimir force coincides, in this case, with its
commutative counterpart. The same happens when $LL$ conditions are imposed. 

The situation changes, however, if one consideres  a situation
involving $L$ and $R$ terms. Indeed, using $R$ conditions at $x_2=l$ and
$L$ conditions at $x_2=0$, in the Dirichlet limit we obtain:
the conditions $\widetilde{\phi}(k_0,k_1;l -\frac{\theta k_1}{2})=0$
and  $\widetilde{\phi}(k_0,k_1;\frac{\theta k_1}{2})=0$. The modes are
quantized, but now they satisfy:
\begin{equation}\label{eq:k2}
	k_2= \frac{n \pi}{|l-\theta k_1|} \;,\;\; n \in {\mathbb N}\;.
\end{equation}
This is of course quite different from the previous case, since we see
that, depending on $k_1$, the $k_2$ allowed values correspond to an
effective size $l -\frac{\theta k_1}{2}$. This is again consistent with the
interpretation that the mirrors are `displaced' by a momentum-dependent
amount. In the hybrid case, the displacement is symmetrical about $x_2 =
l/2$, thus it produces an observable effect (unlike what happens in the
$RR$ and $LL$ cases). Note that the hybrid case is more symmetrical than
those cases, since there is a parity symmetry, in the boundary interaction term,
with respect to the middle point $x_2= l/2$.

The Casimir energy per unit length ${\mathcal E}$ (i.e., tension)
corresponding to this case is straightforwardly evaluated, the result
being:
\begin{equation}
{\mathcal E}\;=\; \frac{1}{2} \, \int \frac{dk_0 dk_1}{(2\pi)^2}
\; \ln\Big[ 1 \, - \, e^{ - 2 \sqrt{k_0^2 + k_1^2}  |l - \theta k_1| }\Big]
\;.
\end{equation}
This tension tends to its proper commutative limit when $l^2 >> \theta$.
However, in the opposite (short distance) limit it grows faster  than $l^{-2}$.

We can calculate the free propagator (corresponding to $S_0$) for
the Dirichlet limit:
\begin{eqnarray}\label{eq:twoprop}
	\Delta(k_\parallel;x_2,y_2) &=& \frac{1}{2 k_\parallel} \Big[ e^{-
	k_\parallel |x_2 - y_2|} \nonumber\\
	&-& \frac{1}{1 - e^{-2 k_\parallel |l - \theta k_1|}} \, e^{-
	k_\parallel \big( |x_2 - l + \frac{\theta k_1}{2}| +  |y_2 - l +
	\frac{\theta k_1}{2}| \big) } \nonumber\\ 
	&-& \frac{1}{1 - e^{-2 k_\parallel |l - \theta k_1|}} \, e^{-
	k_\parallel \big( |x_2 - \frac{\theta k_1}{2}| +  |y_2 - 
	\frac{\theta k_1}{2}| \big) } \nonumber\\ 
	&+& \frac{e^{- k_\parallel |l - \theta k_1|}}{1 - e^{-2 k_\parallel
	|l - \theta k_1|}} \, e^{-
	k_\parallel \big( |x_2 - l + \frac{\theta k_1}{2}| +  |y_2 - 
	\frac{\theta k_1}{2}| \big) } \nonumber\\ 
	&+& \frac{e^{- k_\parallel |l - \theta k_1}|}{1 - e^{-2 k_\parallel |l - \theta k_1|}} \, e^{-
	k_\parallel \big( |x_2 - \frac{\theta k_1}{2}| +  |y_2 - l +
	\frac{\theta k_1}{2}| \big) } \Big] \;. 
\end{eqnarray}
Finally, note that there is no impediment to include both kinds of boundary
term at the same point. The result of this procedure amounts to, in the
Dirichlet limit, imposing $\langle 0 |\phi |\epsilon \rangle = \langle
\epsilon |\phi | 0 \rangle = 0$, $\forall \epsilon$. The corresponding
propagator, for the case of two mirrors at $x_2=0$,  may be obtained by
setting $l=0$ above:
\begin{eqnarray}\label{eq:doubleprop}
	\Delta(k_\parallel;x_2,y_2) &=& \frac{1}{2 k_\parallel} \Big[ e^{-
	k_\parallel |x_2 - y_2|} 
	\;-\; e^{- k_\parallel \big( |x_2 + \frac{\theta k_1}{2}| +  |y_2
+ \frac{\theta k_1}{2}| \big) } \nonumber\\
	&-& e^{- k_\parallel \big( |x_2 - \frac{\theta k_1}{2}| +  |y_2 - 
	\frac{\theta k_1}{2}| \big) } \Big] \;. 
\end{eqnarray}
Besides, it is not difficult to see that when a wave packet insides on this kind of mirror, there is an effective `widening' (depending on $k_1$) as a result
of the emergence momentum dependent boundary condition to the left and right of $x_2=0$.
Note that, imposing both conditions yields a more symmetrical situation, in the sense 
that the system is even under parity and under charge conjugation:
 $\phi \to \phi^\dagger$, $\phi^\dagger \to \phi$.

If one rederives the allowed spectra for two of these mirrors, one finds again (\ref{eq:k2}), plus a condition on $k_1$:
\begin{equation}
k_1 = \frac{l}{\theta} \, r  
\end{equation}
where $r$ is an arbitrary rational number.
\subsection{Discrete spectrum}\label{ssec:discretspec}
In this case, we use instead a discrete basis $\{ |n\rangle\}$: 
\begin{equation}
{\mathcal F}_{\mathcal C} \, |n \rangle \;=\; 
\epsilon_n  \, |n\rangle \;,\;\;\; \langle n|m \rangle \,=\, \delta_{nm} \;.
\end{equation}
Expanding the field in this basis:
\begin{equation}
\phi(\tau) = \sum_{nm} \, \phi_{nm}(\tau) |n \rangle \langle m| 
\;,\;\;\;
\phi^\dagger(\tau) = \sum_{nm} \, \phi^*_{mn}(\tau) \,
|n \rangle \langle m| \;.
\end{equation}
In order to give meaning to the discrete case, getting a non-vanishing
result that moreover is similar to the continuous case, we use, in
this case, a Kronecker $\delta$ (rather that a Dirac one) such that:
\begin{equation}
S_b^{(R)} \;=\; 2 \pi \theta \, \lambda_R \, \int d\tau \; \sum_n \,
|\phi_{0n}(\tau)|^2 \;, 
\end{equation}
which in the Dirichlet ($\lambda_R \to \infty$) limit imposes the condition 
\begin{equation}
\phi_{0n}(\tau) \;=\; \langle 0 |\phi(\tau)| n\rangle =
0 \;,\;\; \forall n \;. 
\end{equation}
So we take this kind of condition as the starting point of our derivation
in the discrete spectrum. 
Note that, in order for the condition on the field to be non-empty, one
should make sure that $0$ belongs to the spectrum of ${\mathcal
F}_{\mathcal C}$. Besides, the normalization of the $\delta$ function loses
its meaning here since one has a sum rather than integral. It is quite natural
to use, as the analogue continuous case, the following:
\begin{equation}
\delta_{\mathcal C} \;=\; {\mathcal P}_{\mathcal C} 
\end{equation} 
where ${\mathcal P}_{\mathcal C}$ is the orthogonal projector on the null
space of the operator ${\mathcal F}_{\mathcal C}$. In the notation used
above: ${\mathcal P}_{\mathcal C} = |0\rangle \langle 0|$.

The example we shall produce for this case will be the one of a circle of
radious $R$, with ${\mathcal F}_{\mathcal C} = \delta\big(\sqrt{x_1^2 +
x_2^2} -R\big)$. Since $[x_1,x_2] = i \theta$, one introduces
\mbox{$a=(x_1+i x_2)/\sqrt{2 \theta}$} and
\mbox{$a^\dagger=(x_1-i x_2)/\sqrt{2 \theta}$}. Then:
\begin{equation}
{\mathcal F}_{\mathcal C} \,=\, \sqrt{2 \theta} \, \big( \sqrt{n + 1/2} -
\frac{R}{\sqrt{2 \theta}} \big) \;\;,\;\;\;\; n = a^\dagger a \;.
\end{equation} 

Besides, $R$ is chosen in order to ensure $0$ belong to the spectrum of
${\mathcal F}_{\mathcal C}$. Thus:
\begin{equation}
\frac{R}{\sqrt{2 \theta}} \;=\; \sqrt{ N + \frac{1}{2}} \;\;,\;\;\; N =
0,1,\ldots 
\end{equation}
As in the continuous case, one can impose both $R$ and $L$ conditions
simultaneously.
\subsection{Symmetries}
We have, in the previous cases, expanded the operators in a basis which
diagonalizes the $\delta$-function; in other words, a basis which is
consistent with the symmetries that survive the imposition of the boundary
conditions. A particularly interesting case arises when the same basis also
diagonalizes the $S_f$ term, since in such a case the free propagator (and,
as a by-product, the energies) can be exactly found.  However, the only two
known cases in which $S_f$ can be exactly diagonalized correspond to 
free fields or to a field in the presence of a `critical magnetic field',
which renders the model self-dual under the Langmann-Szabo
duality~\cite{Langmann:2002cc}, or to the presence of
Grosse-Wulkenhaar~\cite{grosse-wullkenhar} confining potentials. The
basis which renders the propagator diagonal is the so-called `matrix base',
and corresponds to harmonic oscillator like states created by
\mbox{$a^\dagger \equiv \frac{x_1 - i x_2}{\sqrt{2 \theta}}$}.  Thus, the
only non-trivial case where the two symmetries agree, correspond to the
circular boundaries, whose radii are determined by the discrete eigenvalue
of the  number operator $a^\dagger a$. They are considered in the following
section.

\section{Self-dual models}\label{sec:self-dual} 
We have seen that the only system where one can find a basis that
simultaneously diagonalizes $S_f$ and $S_b$ consists of a Langmann-Szabo
self-dual action and  a circular
boundary.  To construct self-dual models one can introduce a coupling
between $\phi$ and a critical magnetic field, or to use a
Grosse-Wulkenhaar confining potential term.  In the former, one uses one of two
constant magnetic fields, $B_L$ and $B_R$, corresponding to fundamental and
anti-fundamental $U_\star(1)$ gauge transformation properties, respectively.
The covariant derivatives act on $\phi$ as follows:
\begin{eqnarray}\label{eq:covader} D_\mu^{(L)} \phi &=& \partial_\mu \phi +
i A_\mu^{(L)} \star \phi \nonumber\\ D_\mu^{(R)} \phi &=& \partial_\mu \phi
+ i \phi \star A_\mu^{(R)} \;, \end{eqnarray} with $A_j^{(L,R)} \equiv
-\frac{B^{(L,R)}}{2} \epsilon_{jk}x_k$, and $A_0^{(L,R)} \equiv 0$.  The
resulting `free' action $S_f$ corresponding to each case is:
\begin{eqnarray} 
S^{(R)}_f &=& \int d^3 x \, \left[ \big( D_\mu^{(R)} \phi \big)^\dagger \star
D_\mu^{(R)}\phi + m^2 \phi^\dagger \star \phi \right] \nonumber\\
S^{(L)}_f &=& \int d^3 x \, \left[ \big( D_\mu^{(L)} \phi \big)^\dagger \star
D_\mu^{(L)}\phi + m^2 \phi^\dagger \star \phi \right] \;.
\end{eqnarray}
Self-duality can be achieved under different conditions, depending on
whether $\theta B^{(R)} = - 2$,  $\theta B^{(L)} =  2$, the
resulting free actions being:
\begin{eqnarray}
	S_f^{(R)} &=& \int d^3 x \, \phi^\dagger \star\big[ -\partial_\tau^2 
	+ \frac{2}{\theta} \, (a^\dagger \star a + \frac{1}{2} ) + m^2\big]
	\star \phi
	\nonumber\\
	S_f^{(L)} &=& \int d^3 x \, \phi \star \big[ -\partial_\tau^2 
	+ \frac{2}{\theta} \, (a^\dagger \star a + \frac{1}{2} ) + m^2
	\big] \star \phi^\dagger \;.
\end{eqnarray}
The confining potential case, on the other hand, corresponds to an action
$S_f^{(A)}$:
\begin{eqnarray}
	S_f^{(A)} &=& \int d^3x \big[ (\partial_\mu \phi)^\dagger \star
	\partial_\mu \phi + \frac{2}{\theta^2} \, \phi^\dagger \star x_j
	\star \phi \star x_j + m^2 \phi^\dagger \star \phi\big] \nonumber\\ 
	 &=& \int d^3x \big\{ \phi^\dagger \star (-\partial_\tau^2 + m^2)
	 \phi \nonumber\\
	 &+&\frac{2}{\theta} \big[\phi^\dagger  \star (a^\dagger \star a +
	 \frac{1}{2} ) \phi + \phi \star (a^\dagger \star a + \frac{1}{2}
	 ) \phi^\dagger \big] \big\} \;.
\end{eqnarray}

The most symmetrical action~\cite{oneloop} corresponds to the confining
potential case,
since it has the symmetry under interchange of the field with its adjoint. 
We see here that the there is a natural choice of boundary term: it has to
be the sum of $R$ and $L$, with identical coupling constants.

We then write, in the matrix base, the $S_b$ action
$S_b=S_b^{(R)}+S_b^{(L)}$  for this choice, with $|N
\rangle$ denoting the state corresponding to the radius of ${\mathcal C}$: 
\begin{eqnarray}
S_b^{(R)} &=& 2 \pi \theta \lambda_R \int_\tau \phi^*_{Nn}(\tau)
\phi_{Nn}(\tau) \nonumber\\ 
S_b^{(L)} &=& 2 \pi \theta \lambda_L \int_\tau \phi^*_{nN}(\tau)
\phi_{nN}(\tau) \;. 
\end{eqnarray}
The propagator (in frequency space) can be found exactly:
\begin{eqnarray}
\langle \phi^*_{jk}(\omega) \phi_{nl} (\omega)\rangle&=& 
(2 \pi \theta)^{-1} \; \frac{\delta_{jn} \delta_{kl}}{\omega^2 + m^2 +
\frac{2}{\theta} (j + k + 1) + \lambda (\delta_{jN} + \delta_{kN})} \;.
\end{eqnarray}
In the Dirichlet limit, one can simply solve the resulting constraints, by
restricting the fields to be of the form:
\begin{equation}
\phi(\tau) = \sum_{n, m \neq N} \, \phi_{nm}(\tau) |n \rangle \langle m| 
\;,\;\;\;
\phi^\dagger(\tau) = \sum_{n, m \neq N} \, \phi^*_{mn}(\tau) \,
|n \rangle \langle m| \;.
\end{equation}
This expansion, when inserted into each one of the interaction terms
$S_I^{(1,2)}$ may be reinterpreted as having consistently eliminated one of the basis
elements, $|N\rangle$. 

By the same token, the Casimir energy in this case is just the vacuum energy
for the $\phi_{NN}$ field mode. This object has the same action as a single
oscillator, thus its vacuum energy is: $E = \sqrt{\frac{2}{\theta}} \, ( 2 N
+ 1)$. In terms of $R$, this corresponds to a tension:
\begin{equation}
{\mathcal E} = \frac{1}{2\pi} \, \frac{R}{\theta^{3/2}} \;.
\end{equation}

\section{Conclusions}\label{sec:conc}
We have studied different ways to impose boundary conditions on a
noncommutative QFT. We have seen that, for the case of a complex field,
there are in principle two inequivalent terms which, however, have the same
commutative limit. 
Those terms introduce qualitatively different boundary conditions, and respect
different symmetries. We have shown that they may be interpreted producing
(under certain circumstances) a `widening' of the curve, such that that effect
is compatible with the uncertainty principle for the coordinates.
We also considered self-dual models, showing that in the
most symmetrical situation, imposing Dirichlet conditions is equivalent to
discarding one of the elements from the matrix base. Namely, to 
reducing the space of field configurations. That reduction can be consistently
implemented even when there are interactions, something that does not happen in
for non-symmetric boundary terms. 

\section*{Acknowledgements} C.\ D.\ F.\ acknowledges support from CONICET,
CNEA, ANPCyT and UNCuyo (Argentina). P.\ S.\ has been supported by
a Petroenergy SA - Trafigura studentship at Instituto Balseiro, UNCuyo.

 
\end{document}